\documentclass{llncs}
\usepackage{wrapfig}
\usepackage{etex}
\usepackage[section]{placeins}
\usepackage[english]{babel} 
\usepackage[latin1]{inputenc} 
\usepackage{latexsym}
\usepackage{epsfig} 
\usepackage{listings}
\usepackage{url,xspace} 
\usepackage[draft]{fixme}
\usepackage{color} 
\usepackage{multirow}
\usepackage{latexsym,amssymb}
\usepackage{fancybox,epic}
\usepackage[noend]{algpseudocode}
\usepackage{amsmath}
\usepackage{graphics}
\usepackage{graphicx}
\usepackage{todonotes}
\usepackage{hyperref}

\usepackage{graphicx}
\usepackage{xcolor}
\definecolor[named]{ACMBlue}{cmyk}{1,0.1,0,0.1}
\definecolor[named]{ACMYellow}{cmyk}{0,0.16,1,0}
\definecolor[named]{ACMOrange}{cmyk}{0,0.42,1,0.01}
\definecolor[named]{ACMRed}{cmyk}{0,0.90,0.86,0}
\definecolor[named]{ACMLightBlue}{cmyk}{0.49,0.01,0,0}
\definecolor[named]{ACMGreen}{cmyk}{0.20,0,1,0.19}
\definecolor[named]{ACMPurple}{cmyk}{0.55,1,0,0.15}
\definecolor[named]{ACMDarkBlue}{cmyk}{1,0.58,0,0.21}
\hypersetup{colorlinks,
  linkcolor=ACMDarkBlue,
  citecolor=ACMPurple,
  urlcolor=ACMDarkBlue,
  filecolor=ACMDarkBlue}

\usepackage{epsfig}

\usepackage{booktabs}
 \usepackage[all]{xy}

\usepackage{amssymb}
\usepackage{latexsym}
\usepackage{fancybox,epic}
\usepackage{algorithm}
\usepackage{amsmath}
\usepackage{colortbl}

\usepackage{pgf}
\usepackage{tikz}
\usepackage{amsmath}

\let\subparagraph\paragraph

\usepackage{listings}

\definecolor{deepblue}{rgb}{0,0,0.5}
\definecolor{deepred}{rgb}{0.6,0,0}
\definecolor{deepgreen}{rgb}{0,0.5,0}
\definecolor{halfgray}{gray}{0.55}
\definecolor{ipythonframe}{RGB}{207, 207, 207}

\definecolor{ckeyword}{HTML}{7F0055}
\definecolor{ccomment}{HTML}{3F7F5F}
\definecolor{cnumber}{HTML}{2A0099}
\definecolor{pblue}{rgb}{0.13,0.13,1}
\definecolor{pgreen}{rgb}{0,0.5,0}
\definecolor{pred}{rgb}{0.9,0,0}
\definecolor{pgrey}{rgb}{0.46,0.45,0.48}

\definecolor[named]{ACMBlue}{cmyk}{1,0.1,0,0.1}
\definecolor[named]{ACMYellow}{cmyk}{0,0.16,1,0}
\definecolor[named]{ACMOrange}{cmyk}{0,0.42,1,0.01}
\definecolor[named]{ACMRed}{cmyk}{0,0.90,0.86,0}
\definecolor[named]{ACMLightBlue}{cmyk}{0.49,0.01,0,0}
\definecolor[named]{ACMGreen}{cmyk}{0.20,0,1,0.19}
\definecolor[named]{ACMPurple}{cmyk}{0.55,1,0,0.15}
\definecolor[named]{ACMDarkBlue}{cmyk}{1,0.58,0,0.21}

\lstdefinelanguage{Solidity} {
  keywords={typeof, modifier, function, public, returns, external,
  contract, new, true, false, private, catch, function, return, null, throw, catch, switch, var, if, in, while, do, else, case, break},
  ndkeywords={bool, address, mapping, uint, bytes32, string},
  identifierstyle=\color{black},
  sensitive=false,
  comment=[l]{//},
  morecomment=[s]{/*}{*/},
  commentstyle=\color{ccomment}\ttfamily,
  string=[b]",
  showstringspaces=false,
  morestring=[b]',
  showspaces=false,
  showtabs=false,
  breaklines=true,
  morekeywords={function, contract, returns, return},
  breakatwhitespace=true,
  lineskip=-0.6pt,
  basewidth={0.54em, 0.4em},
  basicstyle=\footnotesize\ttfamily,
  keywordstyle={\color{ckeyword}\scriptsize\bfseries},
  ndkeywordstyle={\color{black}\small\bf},
  commentstyle={\color{ccomment}\itshape},
  stringstyle={\color{pgreen}},
  numberstyle={\scriptsize\color{cnumber}\ttfamily},
  moredelim=[il][\textcolor{pgrey}]{$$},
  moredelim=[is][\textcolor{pgrey}]{\%\%}{\%\%},
}

\newcommand{\scode}[1]{\lstinline[language=Solidity,basicstyle=\small\ttfamily]{#1}}
\newcommand{\code}[1]{\scode{#1}}

\definecolor{shadecolor}{gray}{1.00}
\definecolor{ddarkgray}{gray}{0.75}
\definecolor{darkgray}{gray}{0.30}
\definecolor{light-gray}{gray}{0.87}

\newcommand{\ie}{\emph{i.e.}\xspace}

\newcommand{\eg}{\emph{e.g.}\xspace}

\newcommand{\etal}{\emph{et~al.}\xspace}

\newcommand{\wrt}{\emph{wrt.}\xspace}

\newcommand{\tname}[1]{\textsc{#1}\xspace}

\newcommand{\plname}[1]{\textsf{#1}\xspace}

\newcommand{\toolname}{\tname{\sc Gastap}} 

\newcommand{\solidity}{\plname{Solidity}}
\newcommand{\vyper}{\plname{Vyper}}
\newcommand{\oyente}{\tname{Oyente}}
\newcommand{\oyentestar}{\tname{Oyente*}}
\newcommand{\ethir}{\tname{EthIR}}
\newcommand{\ethirstar}{\tname{EthIR*}}
\newcommand{\SACO}{\tname{Saco}}
\newcommand{\PUBS}{\tname{Pubs}}
\newcommand{\COFLOCO}{\tname{Cofloco}}

\newcommand{\lst}[1]{\lstinline!#1!}

\newcommand{\Get}[1]{\ensuremath{\mbox{\bf \lstinline!get!}}\xspace}

\newcommand{\returnval}[1]{\mbox{\lstinline!return #1!}\xspace}

\newcommand{\extend}[1]{S}

\newcommand{\sot}[1]{\xoverline{s_{#1}}}

\newcommand{\lvot}[0]{\xoverline{lv}}
\newcommand{\svot}[0]{\xoverline{sv}}
\newcommand{\blot}[0]{\xoverline{bc}}

\newcommand{\svar}[1]{s_{#1}}
\newcommand{\spvar}[1]{s'_{#1}}

\newcommand{\gvar}[1]{g_{#1}}
\newcommand{\lvar}[1]{l_{#1}}

\makeatletter
\newsavebox\myboxA
\newsavebox\myboxB
\newlength\mylenA

\newcommand*\xoverline[2][0.75]{%
    \sbox{\myboxA}{$\m@th#2$}%
    \setbox\myboxB\null
    \ht\myboxB=\ht\myboxA%
    \dp\myboxB=\dp\myboxA%
    \wd\myboxB=#1\wd\myboxA
    \sbox\myboxB{$\m@th\overline{\copy\myboxB}$}
    \setlength\mylenA{\the\wd\myboxA}
    \addtolength\mylenA{-\the\wd\myboxB}%
    \ifdim\wd\myboxB<\wd\myboxA%
       \rlap{\hskip 0.7\mylenA\usebox\myboxB}{\usebox\myboxA}%
    \else
        \hskip -0.5\mylenA\rlap{\usebox\myboxA}{\hskip 0.7\mylenA\usebox\myboxB}%
    \fi}
\makeatother

\lstdefinestyle{numbers}
{numbers=left, numberstyle=\tiny}

\title{Running on Fumes\thanks{
This work was funded partially by the Spanish MINECO project
TIN2015-69175-C4-2-R and MINECO/FEDER, UE project
TIN2015-69175-C4-3-R, by Spanish MICINN/FEDER, UE projects
RTI2018-094403-B-C31 and RTI2018-094403-B-C33, by the CM project
S2018/TCS-4314 and by the UCM CT27/16-CT28/16 grant.}}
\subtitle{Preventing Out-of-Gas Vulnerabilities in Ethereum Smart Contracts using
Static Resource Analysis}
 \author{Elvira Albert$^1$ \and   Pablo Gordillo$^1$   \and  Albert Rubio$^1$ \and Ilya Sergey$^2$}

  \institute{
  Complutense University of Madrid,  Spain \and
  Yale-NUS College and School of Computing, NUS, Singapore}

\begin{document}
\maketitle


\begin{abstract}
  Gas is a measurement unit of the computational effort that it will
  take to execute every single operation that takes part in the
  Ethereum blockchain platform. Each instruction executed by the
  Ethe\-reum Virtual Machine (EVM) has an associated gas consumption
  specified by Ethereum. If a transaction exceeds the amount of gas
  allotted by the user (known as gas limit), an \emph{out-of-gas}
  exception is raised.  There is a wide family of contract
  vulnerabilities due to \emph{out-of-gas} behaviors.  We report on
  the design and implementation of \toolname, a Gas-Aware Smart
  contracT Analysis Platform, which takes as input a smart contract
  (either in EVM, disassembled EVM, or in \textsf{Solidity} source
  code) and automatically infers gas upper bounds for all its public
  functions. Our bounds ensure that if the gas limit paid by the user
  is higher than our inferred gas bounds, the contract is free of
  out-of-gas vulnerabilities.
\end{abstract}


\section{Introduction}\label{intro}
In the Ethereum consensus protocol, every operation on a replicated
blockchain state, which can be performed in a transactional manner
by executing a \emph{smart contract} code, costs a certain amount of
\emph{gas}~\cite{yellow}, a monetary value in \emph{Ether}, Ethereum's
currency, paid by a transaction-proposing party.
Computations (performed by invoking smart contracts) that require
\emph{more computational or storage resources}, cost more gas than
those that require fewer resources.  As regards storage, the EVM has
three areas where it can store items: the \emph{storage} is where all
\emph{contract state} variables reside, every contract has its own
storage and it is persistent between external function calls
(transactions) and quite expensive to use; the \emph{memory} is used
to hold temporary values, and it is erased between transactions and is
cheaper to use; the \emph{stack} is used to carry out operations 
and it is free to use, but can only hold a limited
amount of values.

%
%
The rationale behind the resource-aware smart contract semantics,
instrumented with gas consumption, is three-fold.
First, paying for gas at the moment of proposing the transaction does
not allow the emitter to waste other parties' (aka \emph{miners})
computational power by requiring them to perform a lot of worthless
intensive work.
Second, gas fees disincentivize users to consume too much of
replicated \emph{storage}, which is a valuable resource in a
blockchain-based consensus system.
Finally, such a semantics puts a cap on the number of computations
that a transaction can execute, hence prevents attacks based on
non-terminating executions (which could otherwise, \eg, make all
miners loop forever).

In general, the gas-aware operational semantics of EVM has introduced novel
challenges \wrt sound static reasoning about resource consumption,
correctness, and security of replicated computations:
(1)\label{ch:a} While the EVM specification~\cite{yellow} provides the
precise gas consumption of the low-level operations, most of the smart
contracts are written in high-level languages, such as
\solidity~\cite{solidity} or \vyper~\cite{vyper}.
The translation of the high-level language constructs to the low-level
ones makes static estimation of runtime gas bounds challenging (as we
will see throughout this paper), and is implemented in an
\emph{ad-hoc} way by state-of-the art compilers, which are only able
to give constant gas bounds, or return $\infty$ otherwise.
(2)\label{ch:b} As noted in~\cite{madmax}, it is discouraged in the
Ethereum safety recommendations~\cite{SafetyWiki} that the gas
consumption of smart contracts depends on the size of the data it
stores (i.e., the \emph{contract state}), as well as on the size of
its functions inputs, or of the current state of the
blockchain. However, according to our experiments, almost 10\% of the
functions we have analyzed do.
The inability to estimate those dependencies, and the lack of analysis
tools, leads to design mistakes, which make a contract unsafe to run
or prone to exploits.
For instance, a contract whose state size exceeds a certain limit, can
be made forever \emph{stuck}, not being able to perform any operation
within a reasonable gas bound. Those vulnerabilities have been
recognized before, but only discovered by means of unsound,
pattern-based analysis~\cite{madmax}.

%
%
In this paper, we address these challenges in a principled way by
developing \toolname, a \emph{Gas-Aware Smart contracT Analysis
  Platform}, which is, to the best of our knowledge, the first
automatic gas analyzer for smart contracts.
\toolname takes as input a smart contract provided in \solidity source
code~\cite{solidity}, or in low-level (possibly
decompiled~\cite{porosity}) EVM code, and automatically infers an
upper bound on the gas consumption for each of its public functions.
The upper bounds that \toolname
infers are given in terms of the sizes of the input parameters of the
functions, the contract state, and/or on the blockchain data that the gas
consumption depends upon (e.g., on the \emph{Ether} value).

The inference of gas requires complex transformation and analysis
processes on the code that include: (1) construction of the
control-flow graphs (CFGs), (2) decompilation from low-level code to a
higher-level representation, (3) inference of size relations, (4)
generation of gas equations, and (5) solving the equations into
closed-form gas bounds.  Therefore, building an automatic gas analyzer
from EVM code requires a daunting implementation effort that has been
possible thanks to the availability of a number of existing
open-source tools that we have succeeded to extend and put together in
the \toolname system. In particular, an extension of the tool
\oyente~\cite{oyente} is used for (1), an improved representation of
\ethir \cite{AlbertGLRS18} is used for (2), an adaptation of the size
analyzer of \SACO \cite{AlbertAFGGMPR14} is used to infer the size
relations, and the \PUBS \cite{pubs} solver for (5).

The most challenging aspect in the design of \toolname has been the
approximation of the EVM gas model (which is formally specified in
\cite{yellow}) that is required to produce the gas equations in step
(4). This is because the EVM gas model is highly complex and
unconventional. The gas consumption of each instruction has two parts:
(i) the \emph{memory gas cost}, if the instruction accesses a location
in memory which is beyond the previously accessed locations (known as
\emph{active} memory \cite{yellow}), it pays a gas proportional to the
distance of the accessed location. (ii) The second part, the
\emph{opcode gas cost}, is related to the bytecode instruction itself.
This component is also complex to infer because it is not always a
constant amount, it might depend in some cases on the current global
and local state. 

\toolname has a wide range of applications for contract developers,
attackers and owners, including the detection of vulnerabilities,
debugging and verification/certification of gas usage.  As contract
developers and owners, having a precise resource analyzer allows
answering the following query about a specific smart contract: ``what
is the amount of gas necessary to \emph{safely} (i.e., without an
out-of-gas exception) reach a certain execution point in the contract
code, or to execute a function''? 
This can be used for debugging, verifying/certifying a safe amount of
gas for running, as well as ensuring progress conditions.  
Besides, \toolname allows us to calculate the safe amount of gas
that one should provide to an external data source (e.g., contracts
using Oraclize\cite{oraclize}) in order to enable a successful
callback.
As an attacker, 
one might estimate, how much \emph{Ether}
(in gas), an adversary has to pour into a contract in order to execute
the DoS attack.
We note that such an attack may, however, be economically impractical.

Finally, we argue that our experimental evaluation shows that
\toolname is an effective and efficient tool: we have analyzed more
than 29,000 real smart contracts pulled from \textsf{etherscan.io}~\cite{etherscanSourceCodes},
that in total contain 258,541 public functions, and inferred gas
bounds for 91.85\% of them in 342.54 hours. \toolname can be used from
a web interface at \url{https://costa.fdi.ucm.es/gastap}.



\section{Description of \toolname Components}

\begin{figure}\vspace{-1cm}
\includegraphics[scale=0.4]{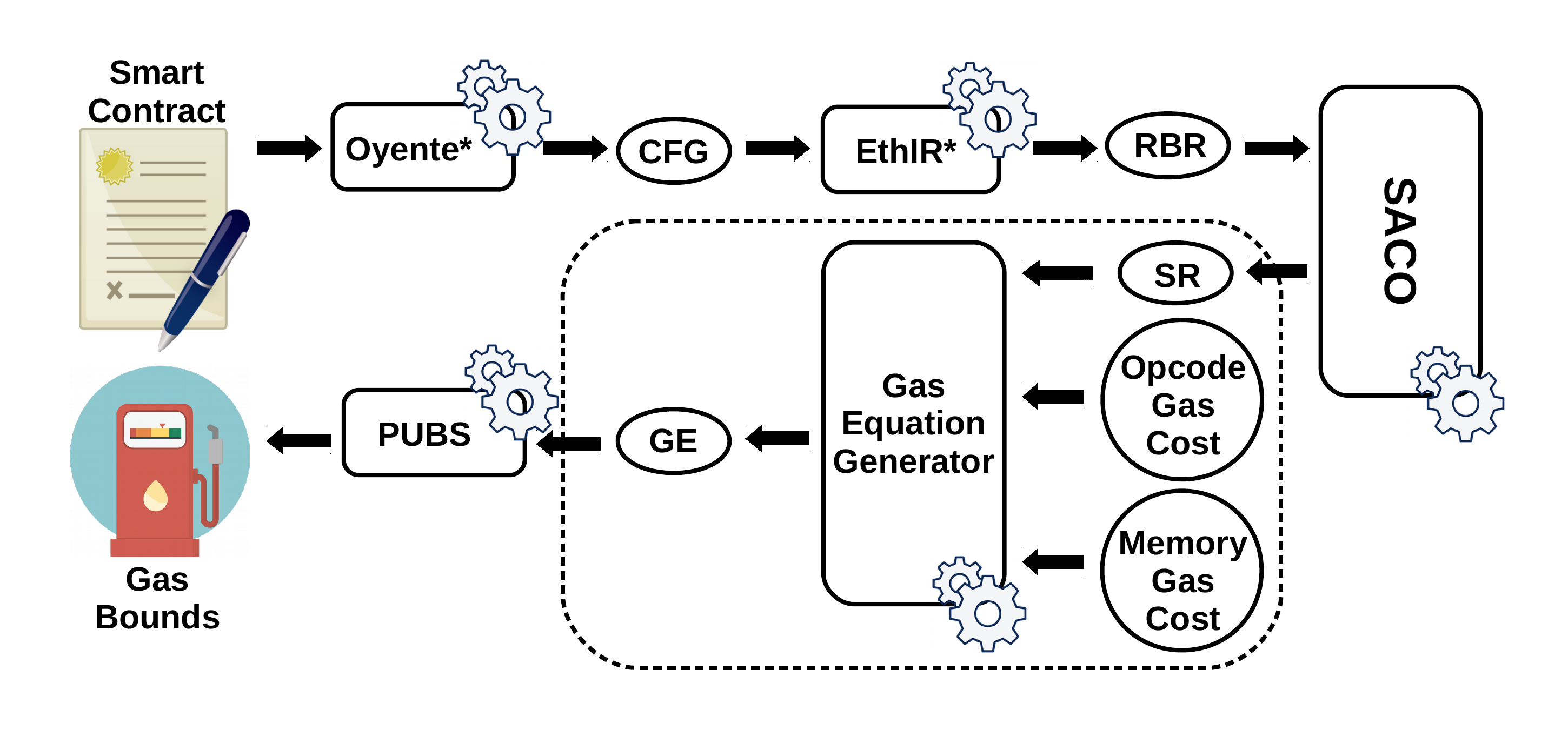} \vspace{-0.65cm}
\caption{Architecture of \toolname (CFG: control flow graph; RBR: rule-based representation;
  SR: size-relations; GE: gas equations)}  \label{tool} 
\end{figure}

Figure~\ref{tool} depicts the architecture of \toolname.
 In order to describe all components of our tool, 
we use as
running example a simplified version (without calls to the external service Oraclize and the authenticity proof verifier)
of the 
\code{EthereumPot} contract~\cite{etherpot} that implements a simple
lottery.
During a game, players 
call a method \code{joinPot} to buy lottery
tickets; each player's address is appended to an array
\code{addresses} of current players, and the number of tickets is
appended to an array \code{slots}, both having variable length.
After some time has elapsed, anyone
can call \code{rewardWinner} which calls the \code{Oraclize}
service to obtain a random number for the winning ticket.  If all goes
according to plan, the \code{Oraclize} service then responds by
calling the \code{__callback} method with this random number and the authenticity proof as arguments.
A new instance of the game is then started, and the winner
is allowed to withdraw her balance using a \code{withdraw} method. In
Fig.~\ref{fig:solevm}, an excerpt of the \textsf{Solidity} code
(including the public  function \code
{findWinner}) and a
fragment of the EVM code produced by the compiler, are displayed. The
 \textsf{Solidity} source code is showed for
readability, as \toolname analyzes directly the EVM code (if it
receives the source, it first compiles it to obtain the EVM code).

\begin{figure}[t]
  \begin{minipage}{0.74\textwidth}
  \begin{center}
  
\begin{tabular}{|l|}
\hline
\hspace{-1pt}
\begin{lstlisting}[name=sol,basicstyle=\scriptsize,keywordstyle=\bfseries\scriptsize]
contract EthereumPot {
 address[] public addresses;
 address public winnerAddress;
 uint[] public slots;
 $\cdots$
 function __callback(bytes32 _queryId, string _result, bytes _proof)
   oraclize_randomDS_proofVerify(_queryId, _result, _proof) {
  if(msg.sender != oraclize_cbAddress()) throw; 
  random_number = uint(sha3(_result))
  winnerAddress = findWinner(random_number);
  amountWon = this.balance * 98 / 100 ;
  winnerAnnounced(winnerAddress, amountWon);
  if(winnerAddress.send(amountWon)) {
   if(owner.send(this.balance)) {
     openPot();
   }}
  }
 function findWinner(uint random) constant returns(address winner){ 
  for(uint i = 0; i < slots.length; i++) {
   if(random <= slots[i]) {
    return addresses[i];
   }}
 }
 $\cdots$ 
}
\end{lstlisting}
\\ 
\hline 
\end{tabular}

%
%
%
%
%

  \end{center}
  \end{minipage}
  \hspace{0.2cm}
  \begin{minipage}{0.23\textwidth}
   \begin{tabular}{|l|}
\hline 
\hspace{3pt}
\begin{lstlisting}[name=sol,basicstyle=\tiny,keywordstyle=\bfseries\scriptsize]
$\cdots$
DUP1
PUSH1 => 0x00
SWAP1
POP
PUSH1 => 0x03
DUP1
SLOAD
SWAP1
$\cdots$
PUSH1 => 0x40
MLOAD
DUP1
SWAP2
SUB
SWAP1
SHA3
PUSH1 => 0x01
$\cdots$
JUMPDEST
MOD
ADD
PUSH1 => 0x0a
DUP2
SWAP1
SSTORE
POP
PUSH2 => 0x0954
PUSH1 => 0x0a
SLOAD 
PUSH2 => 0x064b
JUMP
$\cdots$
\end{lstlisting}
~~\\ 
\hline 
\end{tabular}


  \end{minipage} \vspace{-0.4cm}
  \caption{Excerpt of \textsf{Solidity} code for \code{EthereumPot} contract (left), and \hspace{2cm} $\mbox{ }$ \hspace*{1cm} fragment of EVM code for
 function \code{__callback} (right)} \vspace*{-0.1cm} \vspace{-0.1cm}
\label{fig:solevm}
\end{figure}


\subsection{\oyentestar: from EVM to a complete CFG}\label{sec:oyente}
The first component of our tool, \oyentestar, is an extension of the
open-source tool \oyente~\cite{oyente}, a symbolic
execution tool developed to analyze Ethereum smart contracts and find
potential security bugs.
As \oyente's aim is on symbolic execution rather than on generating a
complete CFG, some extensions are needed to this end. The \ethir
framework~\cite{AlbertGLRS18} had already extended \oyente for two
purposes: (1) to recover the list of addresses for unconditional
blocks with more than one possible jump address (as \oyente originally
only kept the last processed one), and (2) to add more explicit
information to the CFG: jump operations are decorated with the jumping
address,  discovered by \oyente, and, other operations
like store or load are also decorated with the address they operate:
 the number of state variable for operations on
storage; and the memory location for operations on memory if \oyente is able to discover it (or with ``?''
otherwise).

However \ethir's extension still produced incomplete CFGs. \oyentestar
further extends it to handle a more subtle source of incompleteness in
the generated CFG that comes directly from the fact that \oyente is
a symbolic execution engine. For symbolic execution, a bound on the
number of times a loop is iterated is given. Hence it may easily
happen that some (feasible) paths are not reached in the exploration
within this bound and they are lost. 
To solve this problem, we have
modified \oyente to remove the execution bound (as well as other
checks that were only used for their particular applications), and
have added information to the path under analysis. Namely, every time
a new jump is found, we check if the jumping point is already present
in the path. In such case, an edge to that point is added and the
exploration of the trace is
stopped. As a side effect, we not only produce a complete CFG, but
also  avoid much useless exploration for our purposes which results in important 
efficiency gain.


%

 When applying \oyentestar, our extended/modified version of \oyente,
we obtain a \emph{complete} CFG, with the additional annotations
already provided by~\cite{AlbertGLRS18}.





\subsection{\ethirstar: from CFG to an annotated rule-based representation}\label{sec:ethirstar:-from-cfg}

\ethirstar, an extension of \ethir~\cite{AlbertGLRS18}, is the next
component of our analyzer. \ethir provides a rule-based representation
(RBR) for the CFG obtained from \oyentestar. Intuitively, for each
block in the CFG it generates a corresponding rule that contains a
high-level representation of all bytecode instructions in the block
(e.g., load and store operations are represented as assignments) and
that has as parameters an explicit representation of the stack, local,
state, and blockchain variables (details of the transformation are
in~\cite{AlbertGLRS18}). Conditional branching in the CFG is
represented by means of guards in the rules. \ethirstar provides three
extensions to the original version of \ethir~\cite{AlbertGLRS18}:
(1) The first extension is related to the way function calls are
handled in the EVM, where instead of an explicit \texttt{CALL} opcode,
as we have seen before, a call to an internal function is transformed
into a \texttt{PUSH} of the return address in the stack followed by a
\texttt{JUMP} to the address where the code of the function
starts. 
%
If the same function is called from different points of the program,
the resulting CFG shares for all these calls the same subgraph (the
one representing the code of the function) which ends with different
jumping addresses at the end.  As described in~\cite{madmax}, there is
a need to clone parts of the CFG to explicitly link the \texttt{PUSH}
of the return address with the final \texttt{JUMP} to this
address.
%
This cloning in our implementation is done at the level of the RBR as
follows: Since the jumping addresses are known thanks to the symbolic
execution applied by \oyente, we can find the connection between the
\texttt{PUSH} and the \texttt{JUMP} and clone the involved part of the
RBR (between the rule of the \texttt{PUSH} and of the \texttt{JUMP})
using different rule names for each cloning.
%
(2) The second extension is a flow analysis intended to reduce the
number of parameters of the rules of the RBR. This is crucial for
efficiency as the number of involved parameters is a bottleneck for
the successive analysis steps that we are applying. Basically, before
starting the translation phase, we compute the inverse connected
component for each block of the CFG, i.e, the set of its predecessor
blocks. During the generation of each rule, we identify the local,
state or blockchain variables that are used in the body of the
rule. Then, these variables have to be passed as arguments only to
those rules built from the blocks of its inverse connected component.
(3) When we find a store on an unknown memory location
``?'', we have to ``forget'' all the memory from that point on, since
the writing may affect any memory location, and it is not sound
anymore to assume the previous information. In the RBR, we achieve
this deletion by assigning fresh variables (thus unknown values) to
the memory locations at this point.

Optionally, \ethir provides in the RBR the original bytecode
instructions (from which the higher-level ones are obtained) by simply
wrapping them within a nop functor (see Fig. \ref{fig:nop}). Although
nop annotations will be ignored by the size analysis,
they are needed later 
to assign a precise gas consumption to every rule.

\vspace*{-0.5cm}
\begin{figure}[ht]
  \begin{center}
{\scriptsize
\(
\begin{array}{|l|}
  \hline
  \begin{array}{lll}
{\it block1647}(\sot{10}, \svot,\lvot, \blot) \Rightarrow \\
\qquad nop({\tiny JUMPDEST}), \svar{11} = \svar{9}, \svar{9} = \svar{10}, \svar{10} =
\svar{11}, nop(SWAP), \svar{11} = 0, nop(PUSH),\\
\qquad \underline{\lvar{2} = \svar{10}}, nop(MSTORE), 
 \svar{10} = 32, nop(PUSH), 
\svar{11} = 0, nop(PUSH), \underline{\svar{10} = {\it sha3}(\svar{11},
    \svar{10})},\\
\qquad nop(SHA3), \svar{9} = \svar{10}+\svar{9}, nop(ADD),  gl = \svar{9}, 
\svar{9} = fresh_0, nop(SLOAD), \svar{10} = \svar{6},\\
  \qquad  nop(DUP4), {\it call}({\it jump1647}(\sot{10},\svot,\lvot, \blot)),
    nop(GT),nop(ISZERO), nop(ISZERO),\\
\qquad nop(PUSH),nop(JUMPI)\\
  \end{array}\\
  \hline
\end{array}
\)
}
\end{center}
\vspace*{-0.25cm}\caption{Selected rule including nop functions needed
  for gas analysis}\vspace*{-0.7cm}
\label{fig:nop}
\end{figure}

  \begin{example} Figure~\ref{fig:nop} shows the RBR for
    \emph{block1647}. Bytecode instructions that load or
    store information are transformed into assignments on the involved
    variables. For arithmetic operations, operations on bits, sha,
    etc., the variables they operate on are made explicit. Since stack
    variables are always consecutive we denote by $\sot{n}$ the
    decreasing sequence of all $\svar{i}$ form $n$ down to $0$.
    $\lvot$ includes $\lvar{2}$ and $\lvar{0}$, which is the subset of
    the local variables that are needed in this rule or in further
    calls (second extension of \ethirstar). The unknown location ``?''
    has become a fresh variable \emph{fresh$_{0}$} in
    \emph{block1647}. For state variables, $\svot$ includes the needed
    ones
    $\gvar{11},\gvar{8},\gvar{7},\gvar{6},\gvar{5},\gvar{3},\gvar{2},\gvar{1},\gvar{0}$
    ($\gvar{i}$ is the $i$-th state variable). Finally, $\blot$
    includes the needed blockchain state variables \texttt{\small
      address}, \code{balance} and \code{timestamp}.

\end{example}

\subsection{SACO: size relations for EVM smart contracts} \label{size}
In the next step, we generate \emph{size relations} (SR) from the RBR
using the \SACO tool~\cite{AlbertAFGGMPR14}. SR are equations and
inequations that state how the sizes of data change in the rule \cite{DBLP:conf/popl/CousotH78}.
This information is obtained by analyzing how each instruction of the
rules modifies the sizes of the data it uses, and propagating this
information as usual in dataflow analysis.  SR are needed to build
the gas equations and then generate gas bounds in the last step of the
process.  The size analysis of \SACO has been slightly modified to
ignore the $nop$ instructions. Besides, before sending the rules to \SACO, we
replace the instructions that cannot be handled (e.g., bit-wise
operations, hashes) by assignments with fresh variables (to represent
an unknown value).
Apart from this, we are able to adjust our representation to
make use of the approach followed by \SACO, which is based on
abstracting data (structures) to their sizes. For integer variables,
the size abstraction corresponds to their value and thus it works
directly.
However, a
language specific aspect of this step is the handling of data
structures like array, string or bytes (an array of byte).  In the
case of array variables,  \SACO's size analysis works directly as in EVM the
slot assigned to the variable contains indeed its length (and the
address where the array content starts is obtained with the hash of the slot
address).
\vspace{-0.1cm}
\begin{example}\label{size}
  Consider the following SR (those in brackets) generated for rule
  \emph{jump1649} and
  \emph{block1731}:\\
  ${\it jump1619}(\sot{10},\svot,\lvot, \blot) = {\it
    block1633}(\sot{8},\svot,\lvot, \blot) \{\svar{10}<\svar{9}\}$\\
  ${\it block1731}(\sot{8}, \svot,\lvot, \blot) = 41 + {\it
    block1619}(\spvar{8},\sot{7},\svot,\lvot, \blot)
  \{\spvar{8}=1+\svar{8}\}$\\
  The size relations for the \emph{jump1619} function involve the
  \code{slots} array length ($g_3$ stored in $s_9$) and the local
  variable \code{i} (in $s_8$ and copied to $s_{10}$). It corresponds
  to the guard of the \code{for} loop in function \code{findWinner}
  that compares \code{i} and \code{slots.length} and either exits the
  loop or iterates (and hence consume different amount of gas). The
  size relation on $s_8$ for \emph{block1731} corresponds to the size
  increase in the loop counter.
\end{example}
\vspace{-0.1cm}
However, for bytes and
string it is more challenging, as the way they are stored depends on
their actual sizes. Roughly, if they are short (at most $31$ bytes
long) their data is stored in the same slot together with its
length. Otherwise, the slot contains the length (and the address where
the string or bytes content starts is obtained like for arrays). Our
approach to handle this issue is as follows. 
In the presence of bytes or string, we can find in the rules
of the RBR a particular sequence of instructions (which are always the
same) that start pushing the contents of the string or bytes variable in the top of the stack, obtain its length, and leave it stored in the top of the
stack (at the same position). 
Therefore, to avoid losing information, since \SACO is abstracting the
data structures to their sizes, every time we find this pattern of
instructions applied to a string or bytes variable, we just remove them
from the RBR (keeping the nops to account for their gas). Importantly, since the top of the stack has
indeed the size, 
under \SACO's abstraction it is equal to the string or bytes variable. 
Being precise, assuming that we have placed the contents of the string
or bytes variable in the top of the stack, which is $\svar{i}$, the
transformation applied is the following:
$$
{\footnotesize\begin{array}{c}
  \begin{array}{|l|} 
    \hline
  \svar{i+1} = 1, nop(PUSH1), \svar{i+2} = \svar{i}, nop(DUP2),
  \svar{i+3} = 1, nop(PUSH1),\\
  \svar{i+2} = and(\svar{i+3}, \svar{i+2}), nop(AND), 
\svar{i+2}= eq(\svar{i+2}, 0), nop(ISZERO), \\
\svar{i+3} = 256, nop(PUSH2), \svar{i+2} = \svar{i+3}*\svar{i+2}, nop(MUL), \svar{i+1} = \svar{i+2}-\svar{i+1}, ~~~\\
nop(SUB) \svar{i} = and(\svar{i+1}, \svar{i}), nop(AND), \svar{i+1} = 2, nop(PUSH1), \\
\svar{i+2} = \svar{i}, \svar{i} = \svar{i+1}, \svar{i+1} = \svar{i+2}, nop(SWAP1), 
\svar{i} = \svar{i+1}/\svar{i}, nop(DIV) \\
\hline
\end{array} \\
\Downarrow \\
  \begin{array}{|l|}
    \hline 
  nop(PUSH1), nop(DUP2),nop(PUSH1), nop(AND), nop(ISZERO), nop(PUSH2), \\
   nop(MUL), nop(SUB), nop(AND), nop(PUSH1), nop(SWAP1),  nop(DIV) \\
\hline
\end{array}
\end{array}
}$$
Since the involved instructions include
bit-wise operations among others and, as said, the value of the stack variable
becomes unknown, without this transformation the relation between the stack variable and
the length of the string or bytes would be lost and, as a result, the tool
may fail to provide a bound on the gas consumption. This transformation is
 applied when possible and, e.g., is needed to infer bounds
for the functions \code{getPlayers} and \code{getSlots} (see
Table~\ref{fig:large-experiments}).


\subsection{Generation of  equations}\label{sec:gener-gas-equat-1}

In order to generate gas equations (GE), we need to define the EVM gas
model, which is obtained by encoding the specification of the gas
consumption for each EVM instruction as provided in \cite{yellow}. The
EVM gas model is complex and unconventional, it has two components,
one which is related to the memory consumption, and another one that
depends on the bytecode executed. The first component is computed
separately as will be explained below. In this section we focus on
computing the gas attributed to the opcodes. For this purpose, we
provide a function $C_{opcode}:s \mapsto g$ which, for an EVM opcode,
takes a stack $s$ and returns a gas $g$ associated to it. We
distinguish three types of instructions:
(1) Most bytecode instructions have a \emph{fixed} constant gas consumption
  that we encode precisely in the cost model $C_{opcode}$, i.e., $g$
  is a constant.
(2) Bytecode instructions that have different \emph{constant} gas
  consumption $g_1$ or $g_2$ depending on some given condition. This
  is the case of \texttt{SSTORE} that costs $g_1=20000$ if
  the storage value is set from zero to non-zero (first assignment),
  and $g_2=5000$ otherwise. But it is also the case for \texttt{CALL}
  and \texttt{SELFDESTRUCT}. In these cases we use $g=max(g_1,g_2)$ in
  $C_{opcode}$.
(3) Bytecode instructions with a non-constant (\emph{parametric})
  gas consumption that depends on the value of some stack
  location. For instance, the gas consumption of \texttt{EXP} is
  defined as $10+10\cdot(1+\lfloor log_{256}(\mu_{s}[1])\rfloor)$ if
  $\mu_{s}[1]\neq 0$ where $\mu_s[0]$ is the top of the
  stack. Therefore, we have to define $g$ in $C_{opcode}$ as a
  parametric function that uses the involved location. Other bytecode
  instructions with parametric cost are \texttt{CALLDATACOPY},
  \texttt{CODECOPY}, \texttt{RETURNDATACOPY}, \texttt{CALL},
  \texttt{SHA3}, \texttt{LOG*}, and \texttt{EXTCODECOPY}.

Given the RBR annotated with the nop information, the size relations,
and the cost model $C_{opcode}$, we can generate GE that define the
gas consumption of the corresponding code applying the classical
approach to cost analysis \cite{DBLP:journals/cacm/Wegbreit75} which
consists of the following basic steps:
(i) Each rule is transformed into a corresponding cost equation that
defines its cost.  Example~\ref{size} also displays the GE obtained
for the rules \emph{jump1619} and \emph{block1731}.  (ii) The nop
instructions determine the gas that the rule consumes according to the
gas cost model $C_{opcode}$ explained above.  (iii) Calls to other
rules are replaced by calls to the corresponding cost equations. See
for instance the call to \emph{block1619} from rule \emph{block1731}
that is transformed into a call to the cost function \emph{block1619}
in Ex.~\ref{size}. (iv) Size relations are attached to rules to define
their applicability conditions and how the sizes of data change when
the equation is applied. See for instance the size relations attached
to \emph{jump1619} that have been explained in Ex.~\ref{size}.
%

%

As said before, the gas model includes a cost that comes from the
memory consumption which is as follows. Let $C_{mem}(a)$ be the memory
cost function for a given memory slot $a$ and defined as $ 
G_{memory}\cdot a + \left\lfloor{\frac{a^2}{512}}\right\rfloor \mbox{
  where $G_{memory}=3$}$.
Given an EVM instruction, $\mu'_i$ and $\mu_i$
denote resp. the \emph{highest memory slot} accessed in the local memory,
resp., after and before the execution of such instruction.  The
memory gas cost of every instruction is the difference
$C_{mem}(\mu'_i)-C_{mem}(\mu_i)$.
Besides \code{MLOAD} or \code{MSTORE},
instructions like \code{SHA3} or \code{CALL}, among others, make
use of the local memory, and hence can increase the memory gas cost.

In order to estimate this cost associated to all EVM instructions in the
code of the function, we first make the following observations:
(1) Computing the sum of all the 
  memory gas cost amounts to
  computing the memory cost function for the 
  highest memory slot accessed by the instructions of the function
  under analysis. This is because, as seen, $\mu_i$ and $\mu'_i$ refer
  to this position in each operation and hence we pay for all the
  memory up to this point.  (2) This is not a standard memory
  consumption analysis in which one obtains the total amount of memory
  allocated by the function. Instead, in this case, we infer the
  actual value of the highest slot accessed by any operation executed
  in the function.
\vspace{-0.1cm}
\begin{example}
  Let us show how we obtain the memory gas cost for
  \emph{block1647}. In this case, the two instructions in this block
  that cost memory are underlined in Fig.~\ref{fig:nop} and correspond
  to a \code{MSTORE} and \code{SHA3} bytecodes. In this block, both
  bytecodes operate on slot 0 of the memory, and they cost 3 units of
  gas because they only activate up to slot 1 of the
  memory. 
\end{example}




\subsection{PUBS solver: from equations to closed-form bounds}\label{sec:pubsc-solv-from}
The last step of the gas bounds inference is the generation of a
\emph{closed-form gas upper bound}, i.e., a solution for the GE as a
non-recursive expression. As the GE we have generated have the
standard form of cost relations systems, they can be solved using
off-the-shelf solvers, such as \PUBS \cite{pubs} or \COFLOCO
\cite{cofloco}, without requiring any modification. These systems are
able to find polynomial, logarithmic and exponential solutions for
cost relations in a fully automatic way. The gas bounds computed for
all public functions of \textsf{EthereumPot} using \PUBS can be found in
Table~\ref{fig:experiments}, note that they are parametric on
different state variables, input and blockchain data.



\section{Experimental Evaluation}\label{experiments}
This section presents the results of our evaluation
of \toolname. In Sec.~\ref{accuracy}, we evaluate the accuracy of
the gas bounds inferred by \toolname on the \textsf{EthereumPot} by comparing
them with the bounds computed by the \textsf{Solidity}
compiler. 

In Sec.~\ref{statistics}, we evaluate the efficiency and effectiveness
of our tool by analyzing more than 29,000 Ethereum smart contracts. To
obtain these contracts, we pulled from
\textsf{etherscan.io}~\cite{etherscanSourceCodes} all Ethereum
contracts whose source code was available on January 2018. \toolname is
available at \url{https://costa.fdi.ucm.es/gastap}.

\subsection{Gas Bounds for \textsf{EthereumPot} Case Study}\label{accuracy}
Table~\ref{fig:experiments} shows in column \textbf{solc} the gas
bound provided by the \textsf{Solidity} compiler
\textbf{solc}~\cite{solidity}, and in the next two columns the bounds
produced by \toolname for opcode gas and memory gas, respectively,
for all public functions in the contract. 
If we add the gas and memory bounds, it can be observed that, for
those functions with constant gas consumption, we are  as
accurate as \textbf{solc}. Hence, we do not lose precision due to the
use of static analysis.

%
%
For those 6 functions that \textbf{solc} fails to infer constant gas
consumption, it returns $\infty$. For opcode gas, we are able to infer
precise \emph{parametric} bounds for five of them, \code{rewardWinner}
is linear on the size of the first and third state variables ($g1$ and
$g3$ represent resp. the sizes of the arrays \code{addresses} and
\code{slots} in Fig.~\ref{fig:solevm}), \code{getSlots} and
\code{findWinner} on the third, \code{getPlayers} on the first, and
\code{__callback} besides depends on the value of \code{result}
(second function parameter) and \code{proof} (last parameter). It is
important to note that, although the \textsf{Solidity} source code of
some functions (\eg, of \code{getSlots} and \code{getPlayers}) does
not contain loops, they are generated by the compiler and are only
visible at the EVM level. This also happens, for example, when a
function takes a \emph{string} or \emph{bytes} variable as argument.
This shows the need of developing the gas analyzer at the EVM
level.



For \code{joinPot} we cannot ensure that the gas consumption is
finite without embedding information about the blockchain in the
analyzer. This is because \code{joinPot} has a loop:
\code{for (uint i = msg.value; i >= minBetSize; i-= minBetSize)}
\code{\{tickets++;\}}, where \code{minBetSize} is a state variable that is initialized in the
definition line as \code{uint minBetSize = 0.01ether}, and
\code{ether} is the value of the \emph{Ether} at the time of
executing the instruction. This code has indeed several problems. The
first one is that the initialization of the state variable
\code{minBetSize} to the value \code{0.01ether} does not appear in
the EVM code available in the blockchain. This is because this
instruction is executed only once when the contract is created. So our
analyzer cannot find this instruction and the value of
\code{minBetSize} is unknown (and hence no bound can be found).
Besides, the loop indeed does not terminate if \code{minBetSize} in
not strictly greater than zero (which could indeed happen if
\code{ether} would take zero or a negative value). If we add the
initialization instruction, and embed in the analyzer the invariant
that \code{ether}$> 0$ (hence \code{minBetSize} becomes $ > 0$), then
we are able to infer a bound for \code{joinPot}.

\begin{table}[t]
\scriptsize
  \begin{center}
    \begin{tabular}{l|c|c|c|}
      \hline
      \multicolumn{1}{|l|}{\bf function}& \textbf{solc}
      &\textbf{opcode bound \toolname}&\textbf{memory bound \toolname}  \\
      \hline
                              
      \multicolumn{1}{|l|}{\code{totalBet}}& 790 & 775&15\\ \hline
      \multicolumn{1}{|l|}{\code{locked}}& 706 & 691&15\\ \hline
      \multicolumn{1}{|l|}{\code{getEndTime}}& 534 & 519&15\\ \hline
      \multicolumn{1}{|l|}{\code{slots}}& 837 & 822&15\\ \hline
      \multicolumn{1}{|l|}{\code{rewardWinner}}& $\infty$ &
                                                     80391+5057$\cdot$nat(g3)+5057$\cdot$nat(g1)&18\\ \hline
      \multicolumn{1}{|l|}{\code{Kill}}& 30883 & 30874 & 9\\ \hline
      \multicolumn{1}{|l|}{\code{amountWon}}& 438 & 423&15\\ \hline
      \multicolumn{1}{|l|}{\code{getPlayers}}& $\infty$ &
                                                   1373+292$\cdot$nat(g1-1/32)&\\ 
      \multicolumn{1}{|l|}{}                        &
      &+75$\cdot$nat(g1+31/32)&
                                6$\cdot$nat(g1)+24+$\left\lfloor{\frac{(6\cdot nat(g1)+24)^2}{512}}\right\rfloor$\\ \hline
      \multicolumn{1}{|l|}{\code{getSlots}}& $\infty$ &
                                                 1507+250$\cdot$nat(g3-1/32)&\\ 
      \multicolumn{1}{|l|}{}                        &          &+75$\cdot$nat(g3+31/32)&                                 6$\cdot$nat(g3)+24+$\left\lfloor{\frac{(6\cdot nat(g3)+24)^2}{512}}\right\rfloor$\\ \hline
      \multicolumn{1}{|l|}{\code{winnerAddress}}& 750 & 735&15\\ \hline
      \multicolumn{1}{|l|}{\code{\_\_callback}}& $\infty$ &
                                                    229380+3$\cdot$(nat(proof)/32)&
    \\
\multicolumn{1}{|l|}{}&&+103$\cdot$nat(result/32)&\\
\multicolumn{1}{|l|}{}&&+50$\cdot$nat((32-nat(result)))&  max\_error\\
\multicolumn{1}{|l|}{}&&+5836$\cdot$nat(g3)+5057$\cdot$nat(g1) &\\ \hline
      \multicolumn{1}{|l|}{\code{owner}}& 662 & 647 &15\\ \hline
      \multicolumn{1}{|l|}{\code{endTime}}& 460 & 445 &15\\ \hline
      \multicolumn{1}{|l|}{\code{potTime}}& 746 & 731 &15\\ \hline
      \multicolumn{1}{|l|}{\code{potSize}}& 570 & 555 &15\\ \hline
      \multicolumn{1}{|l|}{\code{joinPot}}& $\infty$ & no\_rf &9\\ \hline
      \multicolumn{1}{|l|}{\code{addresses}}& 1116 & 1101&15\\ \hline
      \multicolumn{1}{|l|}{\code{findWinner}}& $\infty$ &
                                                   1555+779$\cdot$nat(g3) &15
      \\ \hline
      \multicolumn{1}{|l|}{\code{random\_number}}& 548 & 533&15\\
\hline
    \end{tabular}
    \caption{Gas bounds  for  \texttt{EthereumPot}. Function
      \texttt{nat} defined as \texttt{nat(l)=max(0,l).}} \vspace{-0.6cm}
    \label{fig:experiments}
  \end{center} 
\end{table}

For \code{__callback} we guarantee that the memory gas is
\emph{finite} but we cannot obtain an upper bound for it, \toolname
yields a \emph{maximization error} which is a consequence of the
information loss due to the soundness requirement described in
extension 3 of Section~\ref{sec:ethirstar:-from-cfg}.  Intuitively,
maximization errors may occur when the analyzer needs to
compose the cost of the different fragments of the code. For the
composition, it needs to maximize (\ie, find the maximal value) the
cost of inner components in their calling contexts (see \cite{pubs}
for details). If the maximization process
involves memory locations that have been ``forgotten'' by \ethirstar
(variables ``?''), 
the upper bound cannot be inferred. Still, if there is no ranking
function error, we know that all loops terminate, thus the memory gas
consumption is finite.


Finally, this transaction is called always with a constant gas limit
of 400,000. This contrasts with the non-constant gas bound obtained
using \toolname. Note that if the gas spent (without including the
\emph{refunds}) goes beyond the gas limit the transaction ends with an
out-of-gas exception. Since the size of $g3$ and $g1$ is the same as
the number of players, from our bound, we can conclude that from 16
players on the contract is in risk of running out-of-gas and get stuck
as the 400,000 gas limit cannot be changed. So using \toolname we can
prevent an out-of-gas vulnerability: the contract should not allow
more than 15 players, or the gas limit must be increased from that
number on.

\subsection{Statistics for Analyzed Contracts}\label{statistics}

Our experimental setup consists on 29,061 contracts
taken from the blockchain as follows. We pulled all Ethereum contracts
from the blockchain as of January 2018, and removed duplicates. This
ended up in 10,796 files (each file often contains several contracts).
We have excluded the files where the decompilation phase fails in any
of the contracts it includes, since in that case we do not get any
information on the whole file. This failure is due to \oyente in 1,230
files, 
which represents a 11.39\% of the total and to \ethir in 829
files, 
which represents a 7.67\% of the total. The failures of \ethir are
mainly due to the cloning mechanism in involved CFGs for which we fail
to find the relation between the jump instruction and the return
address.

After removing these files, our experimental evaluation has been
carried out on the remaining 8,737
files, containing 29,061 contracts. In total we have analyzed 258,541 
public functions (and all auxiliary functions that are used from
them). Experiments have been performed on an Intel Core i7-7700T at
2.9GHz x 8 and 7.7GB of Memory, running Ubuntu 16.04.  \toolname
accepts smart contracts written in versions of \textsf{Solidity} up to
0.4.25 or bytecode for the Ethereum Virtual Machine v1.8.18. The
statistics that we have obtained in number of functions are summarized
in Table~\ref{fig:large-experiments}, and the time taken by the
analyzer in Table~\ref{fig:large-experiments-time}.  The results for
the opcode and memory gas consumption are presented separately.


\begin{table}[t]
{\small
  \setlength{\tabcolsep}{1.5pt}
  \begin{center}
    \begin{tabular}{|l|c|c|c|c|}
      \hline
      {\bf Type of result}&{\bf \#opc}&{\bf \%opc} &{\bf \#mem}&{\bf \%mem}\\ \hline
      Constant gas bound & 223,294 & 86.37\% &225,860& 87.36\%\\
      Parametric gas bound  & 14,167 & 5.48\% &13,312&5.15\%\\
      Time out & 13,140 & 5.08\%& 13,539 &5.24\%\\
      Finite gas bound (maximization error)& 7,095& 2.74\% & 5,830&2.25\%\\
      Termination unknown (ranking function error) & 716 & 0.28\% &0&0\%\\
      Complex control flow (cover point error) & 129 & 0.05\% &0& 0\%\\\hline
      Total number of functions & 258,541 & 100\% &258,541 &100\%\\\hline
    \end{tabular}
      \end{center} 
    \caption{Statistics of gas usage on the analyzed 29,061 smart contracts from
      Ethereum blockchain}\vspace{-0.6cm}
    \label{fig:large-experiments}
}
\end{table}

Let us first discuss the results in Table~\ref{fig:large-experiments}
which aim at showing the effectiveness of \toolname. 
Columns \textbf{\#opc} and \textbf{\#mem} contain number of analyzed
functions for opcode and memory gas, resp., and columns preceded by
\textbf{\%} the percentage they represent. For the analyzed contracts,
we can see that a large number of functions, 86.37\% (resp. 87.36\%),
have a constant opcode (resp. memory) gas consumption. This is as
expected because of the nature of smart contracts, as well as because
of the Ethereum safety recommendations mentioned in
Section~\ref{intro}.
Still, there is a relevant
number of functions 5.48\% (resp. 5.15\%) for which we obtain an opcode (resp. memory)
gas bound that is not constant (and hence are potentially
vulnerable). 
Additionally, 5.08\% of the analyzed functions for opcodes and 5.24\%
for memory reach the timeout (set to 1 minute) due to the further
complexity of solving the equations. 

As the number of analyzed contracts is very large, a manual inspection
of all of them is not possible. Having inspected many of them
and, thanks to the information provided by the \PUBS solver used by
\toolname, we are able to classify the types of errors that have led
to a ``\emph{don't-know}'' answer and which in turn explain the
sources of incompleteness by our analysis:
%
(i) \emph{Maximization error}:
  In many cases, a \emph{maximization error} is a consequence of loss of
  information by the size analysis or by
  the decompilation when the values of memory locations are lost. As
  mentioned, even if we do not produce the gas formula, we know that
  the gas consumption is \emph{finite} (otherwise the system flags a
  ranking function error described below).
(ii) \emph{Ranking function error:} The solver needs to find ranking
  functions to bound the maximum number of iterations of all loops the
  analyzed code might perform. If \toolname fails at this step, it
  outputs a \emph{ranking function error}. 
  Section~\ref{experiments} has described a scenario where we have
  stumbled across this kind of error. We note that number of these
  failures for \textbf{mem} is lower than for \textbf{opcode} because
  when the cost accumulated in a loop is 0, \PUBS does not look for a
  ranking function.
(iii) \emph{Cover point error:} The equations are transformed into
  direct recursive form to be solved \cite{pubs}. If the
  transformation is not feasible, a \emph{cover point error} is
  thrown. This might happen when we have mutually recursive functions,
  but it also happens for nested loops as in non-structured languages.
  This is because they contain jump instructions from the inner loop
  to the outer, and vice versa, and become mutually recursive. A loop
  extraction transformation would solve this problem, and we leave its
  implementation for the future work.

\begin{table}[t] 
{\small
  \setlength{\tabcolsep}{1.5pt}
  \begin{center}
    \begin{tabular}{|l|c|c|c|c|c|c|c|}
      \hline
      {\bf Phase} & {\bf T$_{opcode}$} (s)  &  {\bf T$_{\mathit{mem}}$} (s) &{\bf T$_{\mathit{total}}$} (s)& {\bf \%opc} & {\bf \%mem} &{\bf \%total}\\ \hline
      CFG generation (\oyentestar) & --- & --- & 17,075.55&---&---&1.384\% \\
      RBR generation (\ethirstar)  & --- & --- &  81.37 &---&---&0.006\%  \\
      Size analysis (\SACO)  &---&---&105,732 &---&---&8.57\%\\
      Generation of gas equations  &  141,576   & 125,760 &267,336 & 11.48\%  &10.2\%&21.68\%\\
      Solving gas equation (\PUBS)  & 395,429 & 447,502 &842,931 & 32.06\%  & 36.3\%&68.36\%\\ \hline
      Total time \toolname  & && 1,233,155.92  &&& 100\% \\ \hline
    \end{tabular}
  \end{center}} 
    \caption{Timing breakdown for \toolname on the analyzed 29,061 smart contracts}\vspace{-.6cm}
    \label{fig:large-experiments-time}
\end{table}

%
%
%

%
As regards the efficiency of \toolname, the total analysis time for
all functions is 1,233,155.92 sec (342.54 hours).  Columns \textbf{T} and
\textbf{\%} show, resp., the time in seconds for each phase and the percentage
of the total for each type of gas bound. The first three rows are common for the
inference of the opcode and memory bounds, while equation generation
and solving is separated for opcode and memory. Most of the time is spent 
in solving the GE (68.36\%), which includes some timeouts.
The time taken by
\ethir is negligible, as it is a syntactic transformation process,
while all other parts require semantic reasoning.
All in all, we argue that the statistics from our experimental
evaluation  show the accuracy, effectiveness and efficiency of our
tool. 
Also, the sources of incompleteness point out directions for further
improvements of the tool.




\section{Related Work and Conclusions}

Analysis of Ethereum smart contracts for possible safety violations
and security and vulnerabilities is a popular topic that has received
a lot of attention recently, with numerous tools developed, leveraging
techniques based on symbolic
execution~\cite{Luu-al:CCS16,GrossmanAGMRSZ18,Nikolic-al:Maian,KruppR18,Kalra-al:NDSS18,TsankovDDGBV18},
SMT solving~\cite{Marescotti-al:ISoLA18,Kolluri-al:laws}, and
certified
programming~\cite{Bhargavan-al:PLAS16,Grishchenko-al:POST18,Amani-al:CPP18},
with only a small fraction of them focusing on analyzing gas
consumption.

The \tname{GASPER} tool identifies gas-costly programming
patterns~\cite{ChenLLZ17}, which can be optimized to consume less. For
doing so, it relies on matching specific control-flow patterns, SMT
solvers and symbolic computation, which makes their analysis neither
sound, nor complete.
In a similar vein, the recent work by Grech~\etal ~\cite{madmax}
identifies a number of classes of gas-focused vulnerabilities, and
provides \tname{MadMax}, a static analysis, also working on a
decompiled EVM bytecode, data-combining techniques from flow analysis
together with CFA context-sensitive analysis and modeling of memory
layout.
In its techniques, \tname{MadMax} differs from \toolname, as it
focuses on identifying control- and data-flow patterns inherent for
the gas-related vulnerabilities, thus, working as a bug-finder, rather
than complexity analyzer. Since deriving accurate worst-case
complexity boundaries is not a goal of any of both \tname{GASPER} and
\tname{MadMax}, they are unsuitable for tackling the
challenge~\ref{ch:a}, which we have posed in the introduction.

In a concurrent work, Marescotti~\etal~identified three cases in which
computing gas consumption can help in making Ethereum more efficient:
(a) prevent errors causing contracts get stuck with
\emph{out-of-gas} exception, (b) place the right price on the gas
unit, and (c)~recognize semantically-equivalent smart
contracts~\cite{Marescotti-al:ISoLA18}.
They propose a methodology, based on the notion of the
so-called \emph{gas consumption paths} (GCPs) to estimate the
worst-case gas consumption using techniques from symbolic bounded
model checking~\cite{Biere-al:TACAS99}. Their approach is based on
symbolically enumerating all execution paths and unwinding loops to a
limit.
Instead, using resource analysis, \toolname infers the maximal number
of iterations for loops and generates accurate gas bounds which are
valid for any possible execution of the function and not only for the
unwound paths.
%
Besides, the approach by Marescotti~\etal has not been implemented in
the context of EVM and has not been evaluated on real-world smart
contracts as ours.
\vspace{-0.3cm}
\paragraph{Conclusions.~}
Automated static reasoning about resource consumption is
critical for developing safe and secure blockchain-based replicated
computations, managing billions of dollars worth of virtual currency.
In this work, we employed state-of-the art techniques in resource
analysis, showing that such reasoning is feasible for Ethereum, where
it can be used at scale not only for detecting vulnerabilities, but
also for  verification/certification of existing smart contracts.

\bibliographystyle{plain}




\end{document}